\documentclass[aps,prl,showpacs,twocolumn,superscriptaddress,longbibliography]{revtex4-1}
\usepackage[pdftex]{graphicx}
\usepackage[pdftex,bookmarks,colorlinks,breaklinks]{hyperref} 
\hypersetup{linkcolor=red,citecolor=blue,filecolor=dullmagenta,urlcolor=blue} 

\begin{document}

\title{Microwave Realization of the Gaussian Symplectic Ensemble}
\author{A. Rehemanjiang}
\affiliation{Fachbereich Physik der Philipps-Universit\"at Marburg, D-35032 Marburg, Germany}
\author{M. Allgaier}
\affiliation{Fachbereich Physik der Philipps-Universit\"at Marburg, D-35032 Marburg, Germany}
\affiliation{Integrated Quantum Optics, Applied Physics, University of Paderborn, 33098 Paderborn, Germany}
\author{C. H. Joyner}
\affiliation{School of Mathematical Sciences, Queen Mary University of London, London E1 4NS, United Kingdom}
\author{S. M\"{u}ller}
\affiliation{School of Mathematics, University of Bristol, Bristol BS8 1TW, United Kingdom}
\author{M. Sieber}
\affiliation{School of Mathematics, University of Bristol, Bristol BS8 1TW, United Kingdom}
\author{U. Kuhl}
\affiliation{Fachbereich Physik der Philipps-Universit\"at Marburg, D-35032 Marburg, Germany}
\affiliation{Universit\'e de Nice - Sophia Antipolis, Laboratoire de la Physique de la Mati\`ere Condens\'ee, CNRS, Parc Valrose, 06108 Nice, France}
\author{H.-J. St\"ockmann}
\affiliation{Fachbereich Physik der Philipps-Universit\"at Marburg, D-35032 Marburg, Germany}

\date{\today}

\begin{abstract}
Following an idea by Joyner {\it et al.}~[Europhys. Lett. 107, 50004 (2014)] a microwave graph with an antiunitary symmetry $T$ obeying $T^2=-1$ is realized. The Kramers doublets expected for such systems are clearly identified and can be lifted by a perturbation which breaks the antiunitary symmetry. The observed spectral level spacings distribution of the Kramers doublets is in agreement with the predictions from the Gaussian symplectic ensemble expected for chaotic systems with such a symmetry.
\end{abstract}

\pacs{05.45.Mt}

\maketitle

Random matrix theory has proven to be an extremely powerful tool to describe the spectra of chaotic systems \cite{meh91,cas80,boh84b,haa01b}. For systems with time-reversal symmetry (TRS) and no half-integer spin in particular, there is an abundant number of theoretical, numerical, and experimental studies showing that the universal spectral properties are perfectly well reproduced by the corresponding properties of the Gaussian orthogonal random matrix ensemble (GOE) (see Ref.~\onlinecite{stoe99} for a review).
This is the essence of the famous conjecture by Bohigas, Giannoni and Schmit \cite{boh84b}, which has received strong theoretical support; see, e.g., Refs.~\onlinecite{ber85,sie01,muel09}.
For systems with TRS and half-integer spin and systems with no TRS the Gaussian symplectic ensemble (GSE) and the Gaussian unitary ensemble (GUE), respectively, hold instead. There are three studies of the spectra of systems with broken TRS showing GUE statistics \cite{so95,sto95b,hul04}, all of them applying microwave techniques. For the GSE there is no experimental realization at all up to now. Only by using that a GSE spectrum is obtained by taking only every second level of a GOE spectrum \cite{meh91}, GSE statistics has been experimentally observed in a microwave hyperbola billiard \cite{alt97b}.

In fact, GUE statistics may be observed even in systems without broken TRS if there is a suitable geometrical symmetry. One example is the billiard with threefold rotational symmetry \cite{ley96a} with microwave realizations \cite{dem00b,sch02}. Another example is the constant width billiard \cite{gut07}, again with an experimental realization \cite{die14}.

On the other hand, GOE statistics may be obtained in billiards with a magnetic field if there is an additional reflection symmetry \cite{ber86b}. This is because there exists an antiunitary symmetry that combines time reversal with reflection. To be able to observe GSE statistics in a system without spin requires a similar nonconventional symmetry. What is needed according to Dyson's threefold way \cite{dys62a} is an antiunitary symmetry $T$ with the property that $T^2=-1$. This is sufficient to guarantee GSE statistics if the system is chaotic \cite{sch88}. In addition, it leads to Kramer's degeneracy; i.\,e., the application of $T$ to an energy eigenfunction yields an orthogonal eigenfunction with equal energy. A system with such a symmetry was recently found in the form of a quantum graph \cite{joy14}.

Quantum graphs were introduced by Kottos and Smilansky \cite{kot97a} to study various aspects of quantum chaos. The wave function on a quantum graph satisfies a one-dimensional Schr\"odinger equation on each of the bonds with suitable matching conditions (implying current conservation) at the vertices. Just as for quantum billiards, there is a one-to-one mapping onto the corresponding microwave graph. This analogy has been used in a number of experiments including one on graphs with and without broken TRS \cite{hul04,hul05a,all14a}.

\begin{figure}
  \raisebox{4cm}[0cm][0cm]{(a)}\hspace*{0.5cm}\includegraphics[width=0.8\columnwidth]{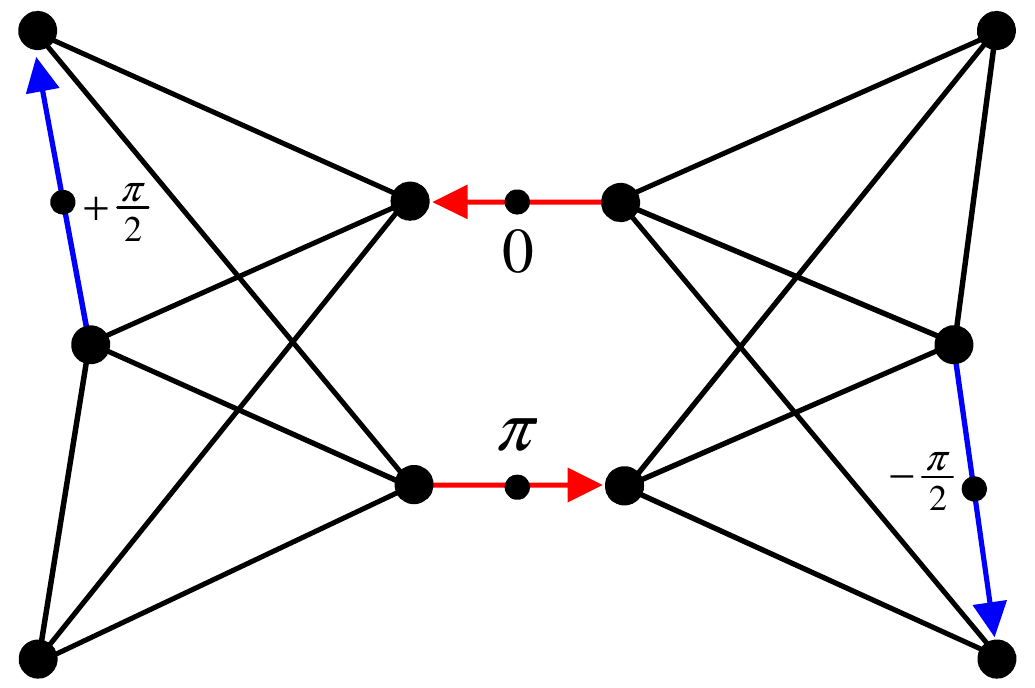}\hspace*{1.cm}\quad\\[1ex]
  \raisebox{4cm}[0cm][0cm]{(b)}\includegraphics[width=0.95\columnwidth]{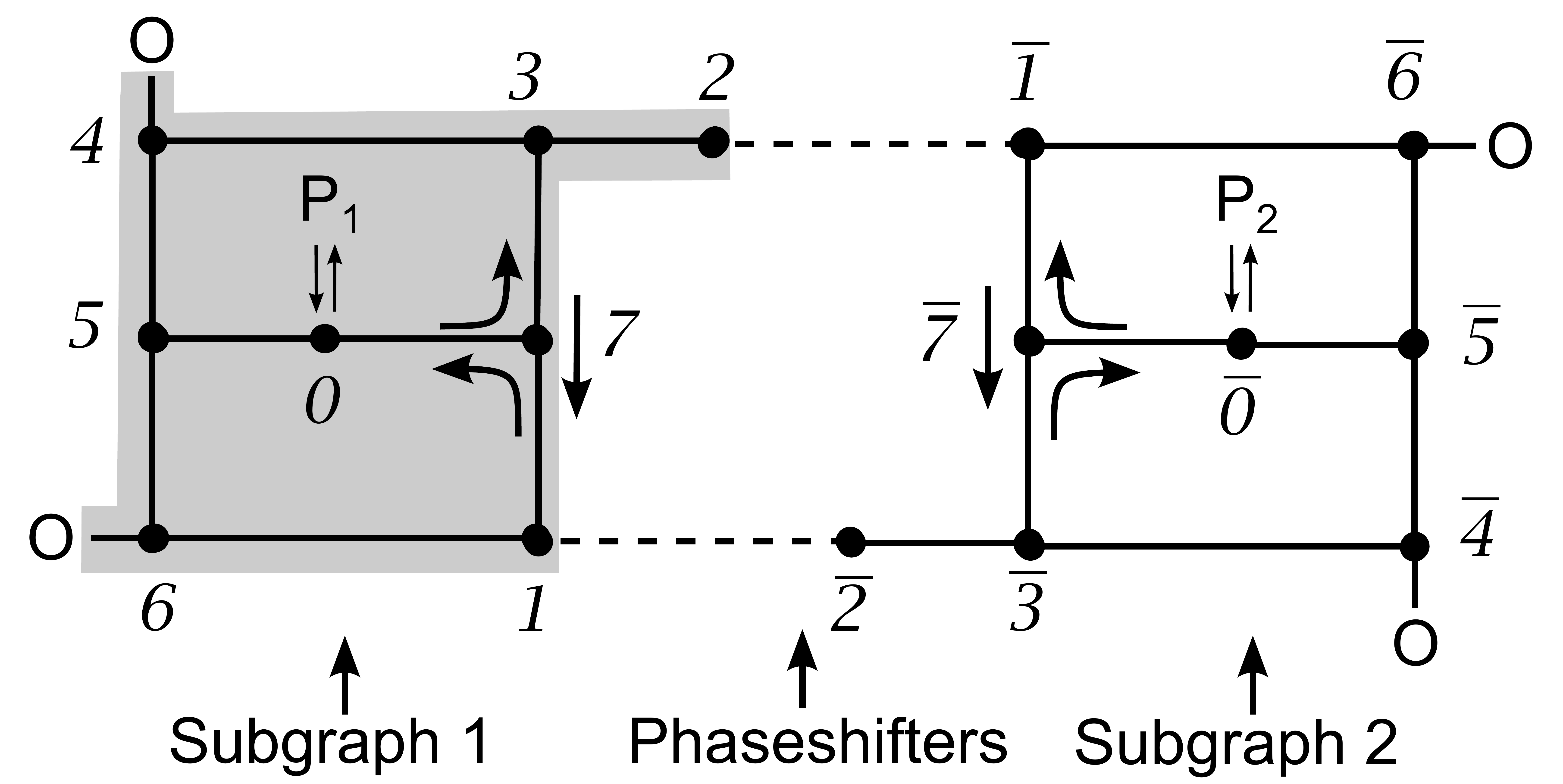}\\[1ex]
  \raisebox{7.5cm}[0cm][0cm]{(c)}\includegraphics[width=0.95\columnwidth]{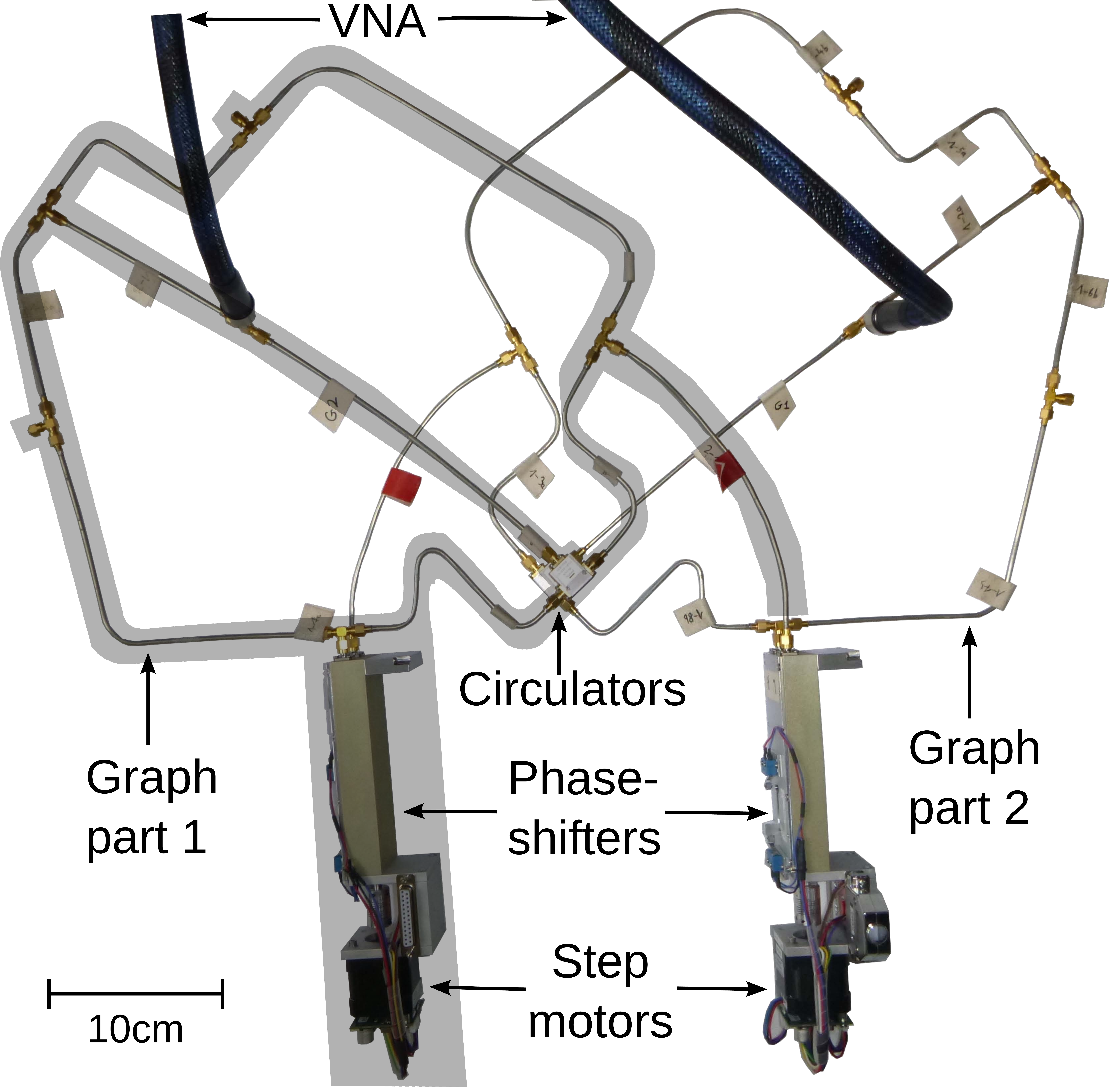}
  \caption{\label{fig:graph} (Color online)
  (a) Sketch of the graph proposed in Ref.~\onlinecite{joy14} to study GSE statistics without spin.
  The four arrows denote bonds along which additional phases are acquired.
  (b) Schematic drawing of one of the realized microwave graphs.
  Subgraph 1 is highlighted by a gray background.
  The dashed lines correspond to phase shifters with variable lengths.
  The two subgraphs contain microwave circulators at nodes $7$ and $\bar{7}$, respectively, with opposite sense of rotation.
  The nodes marked by ``O'' are closed by open end terminators.
  They were used to allow for an easy realization of alternative graphs.
  Subgraphs 1 and 2 are connected at nodes $0$ and $\bar{0}$, respectively, to ports $P_1$ ($P_2$) of the VNA.
  (c) Photograph of the graph sketched in (b) consisting of $T$~junctions, semirigid cables with identification tags, circulators, open end terminators, and phase shifters with step motors.
  Again, subgraph 1 is highlighted.
  }
\end{figure}

To realize graphs with GSE symmetry the graph shown in Fig.~\ref{fig:graph}(top) was proposed in Ref.~\cite{joy14}. It contains two geometrically identical subgraphs but with phase shifts by $+\pi/2$ and $-\pi/2$, respectively, along two corresponding bonds. The two subgraphs are connected by a pair of bonds yielding a graph with a geometric inversion center. In addition, there is another phase shift of $\pi$ along one of the two bonds but not the other one. This is the crucial point: Because of this trick, the total graph is symmetric with respect to an antiunitary operator $T$ squaring to -1, $T^2=-1$, where
\begin{equation} \label{eq:tsymm}
\begin{array}{cc}
T \psi(x_1) & = + \psi^*(x_2) , \\ T \psi(x_2) & = - \psi^*(x_1)\,;
\end{array}
\end{equation}
i.e., if $\psi$ satisfies the Schr\"odinger equation as well as the vertex conditions, then the same applies to $T \psi$.
Here $x_1$ is a coordinate in subgraph~1, and $x_2$ the corresponding coordinate (related by inversion) in subgraph~2. Applying Eq.~(\ref{eq:tsymm})
twice shows $T^2=-1$.

A complementary approach shall be given establishing a direct correspondence between the experiment and a spin 1/2 system. The wave function in a quantum graph is subject to two constraints, continuity at the vertices and current conservation. In a microwave graph these constraints correspond to the well-known Kirchhoff rules governing electric circuits. They lead to a secular equation system having a solution only if the determinant of the corresponding matrix vanishes,
\begin{equation}\label{eq:det}
  \mathrm{det}[h(k)]=0
\end{equation}
where the matrix elements of $h(k)$ are given by
\begin{equation}\label{eq:sec}
  h_{ij}(k)=\left\{
  \begin{array}{cc}
    -\sum\limits_{n\ne i}C_{in}\cot(kl_{in}) & i=j \\
    C_{ij}e^{-{\imath\varphi_{ij}}}\left[\sin(kl_{ij})\right]^{-1} & i\ne j
  \end{array}
  \right.
\end{equation}
where $C_{ij}=1$, if nodes $i$ and $j$ are connected, and $C_{ij}=0$, otherwise. $l_{ij}$ is the length of the bond connecting nodes $i$ and $j$. $\varphi_{ij}$ is a phase resulting, e.\,g., from a vector potential and breaks TRS if present. The equation holds for Neumann boundary conditions at all nodes, the situation found in the experiment. Details can be found in Ref.~\cite{kot99a}. The solutions of the determinant condition (\ref{eq:det}) generate the spectrum of the graph.

Applied to the graph of Fig.~\ref{fig:graph}, the secular matrix $h(k)$ may be written as
\begin{equation}\label{eq:sec1}
  h=h_{\mathrm{dis}}+v
\end{equation}
where $h_{\mathrm{dis}}$ is the secular matrix for the disconnected subgraphs, and $v$ describes the connecting bonds. It is convenient to introduce an order of rows and columns according to $\{1, 2, \dots,n;\bar{1}, \bar{2}, \dots, \bar{n}\}$, where the numbers without a bar refer to the vertices of subgraph 1, and the numbers with a bar to those of subgraph 2. $h_{\mathrm{dis}}$ may then be written as
\begin{equation}\label{eq:sec2}
  h_{\mathrm{dis}}=\left(
  \begin{array}{cc}
    h_0 & \cdot \\
    \cdot & h_0^* \\
  \end{array}
  \right)
\end{equation}
where $h_0$ and $h_0^*$ are the secular matrices for each of the two subgraphs, respectively. Since the only difference between the subgraphs is the sign of the $\pi/2$ phase shift in one of the bonds, their secular matrices are just complex conjugates of each other; see Eq.~(\ref{eq:sec}). Assuming for the sake of simplicity that there is just one pair of bonds connecting node $1$ with node $\bar{2}$, and node $\bar{1}$ with node $2$, the matrix elements of $v$ are given by
\begin{eqnarray}
  v_{11}&=&v_{22}=v_{\bar{1}\bar{1}}=v_{\bar{2}\bar{2}}=-\cot (kl)\,,\\\label{eq:sec3}
  v_{1\bar{2}}&=&v_{\bar{2}1}=-v_{2\bar{1}}=-v_{\bar{1}2}=\left[\sin (kl)\right]^{-1}\,,\\
  v_{ij}&=&v_{\bar{i}\bar{j}}=v_{i\bar{j}}=v_{\bar{i}j}=0\,, \quad \mbox{otherwise,}
\end{eqnarray}
where $l$ is the length of the bond connecting $1$ with $\bar{2}$ and $\bar{1}$ with $2$. The generalization to a larger number of bond pairs is straightforward.

Changing now the sequence of rows and columns to $\{1, \bar{1};2, \bar{2}; \dots; n, \bar{n}\}$, the resulting $2n\times 2n$ matrix $\tilde{h}(k)$ may be written in terms of an $n\times n$ matrix with quaternion matrix elements,
\begin{equation}\label{eq:sec4}
  [\tilde{h}(k)]_{nm}= \left[\mathrm{Re}(h_0)_{nm}+ v_{nm}\right]{\bf 1} -\mathrm{Im}(h_0)_{nm}{\bf\tau_z}-v_{n\bar{m}}{\bf\tau_y}
\end{equation}
where
\begin{equation}\label{eq:sec6}
  {\bf 1}=\left(
  \begin{array}{cc}
    1 & \cdot \\
    \cdot & 1 \\
  \end{array}
  \right),\quad
  {\bf \tau_z}=\left(
  \begin{array}{cc}
    -\imath & \cdot \\
    \cdot & \imath \\
  \end{array}
  \right),\quad
  {\bf \tau_y}=\left(
  \begin{array}{cc}
    \cdot & -1 \\
    1 & \cdot \\
  \end{array}
  \right)
\end{equation}
The determinant is not changed by this rearrangement of rows and columns, $\mathrm{det}[h(k)]=\mathrm{det}[\tilde{h}(k)]$. The matrix elements $[\tilde{h}(k)]_{nm}$ commute with $C{\bf\tau_y}$, where $C$ denotes the complex conjugate, and, hence, the whole matrix commutes with $T=\mathrm{diag}(C{\bf\tau_y}\, \dots,C{\bf\tau_y})$, where $T$ squares to -1, $T^2=-1$. This is exactly the situation found for spin 1/2 systems, and just as in such systems, a twofold Kramers degenerate spectrum is expected showing the signatures of the GSE provided the system is chaotic; see e.g.~Chap.~2 of Ref.~\onlinecite{haa01b}.

The requirements defined by Joyner {\it et\,al.}~\cite{joy14} to realize graphs with GSE symmetry pose some challenges. Since we did not know of a simple way to achieve phase shifts of $\pm\pi/2$ along the bonds, we instead built two geometrically identical subgraphs but with two circulators of opposite sense of rotation within the two subgraphs. A circulator is a T-shaped microwave device introducing directionality: Microwaves pass from port 1 to port 2, from port 2 to port 3, and from port 3 to port 1. The result is the same as with the $\pm\pi/2$ shifts: The circulators break TRS, resulting in identical GUE spectra for the two disconnected subgraphs but with an opposite sense of propagation within the respective subgraphs. Again, the two subgraphs may, thus, be described in terms of a secular matrix $h_0$ and its complex conjugate $h_0^*$, respectively.

The phase difference between the two connecting bonds is adjustable by means of mechanical phase shifters, which in reality, however, do not change the phase but the length. This approach has the shortcoming that for a given length change $\Delta l$, the phase shift $\Delta\varphi$ depends on frequency $\nu$:
\begin{equation}\label{eq:phase}
  \Delta\varphi= k\Delta l= \frac{2\pi\nu}{c}\Delta l
\end{equation}
where $k$ is the wave number, and $c$ is the vacuum velocity of light. $l=nl_0$ is the {\em optical} length where $l_0$ is the {\em geometrical} length and $n=1.43$ the index of refraction of the dielectric within the coaxial cables.

\begin{figure}[b]
  \includegraphics[width=\columnwidth]{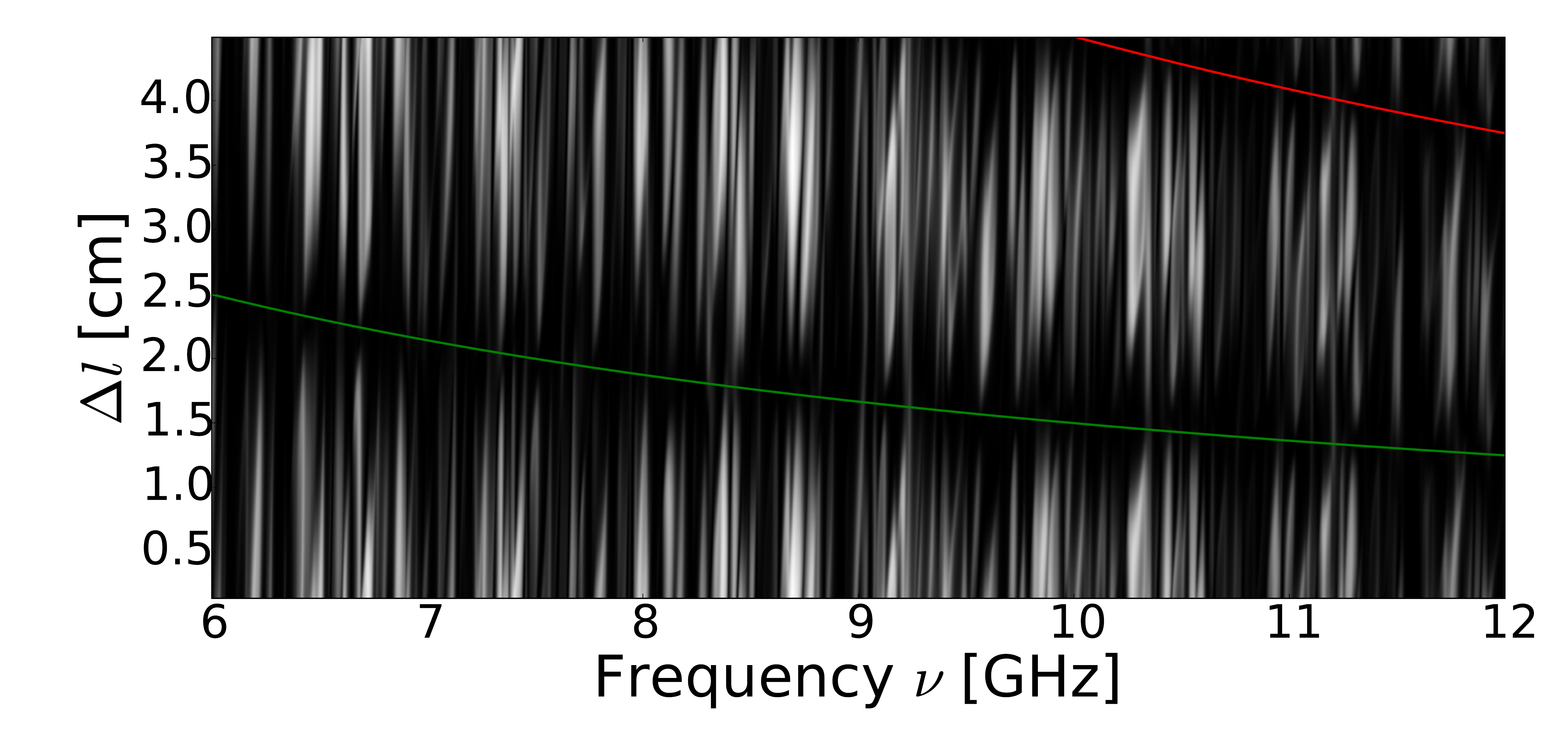}\\[-1.5ex]
  \includegraphics[width=\columnwidth,height=5 cm]{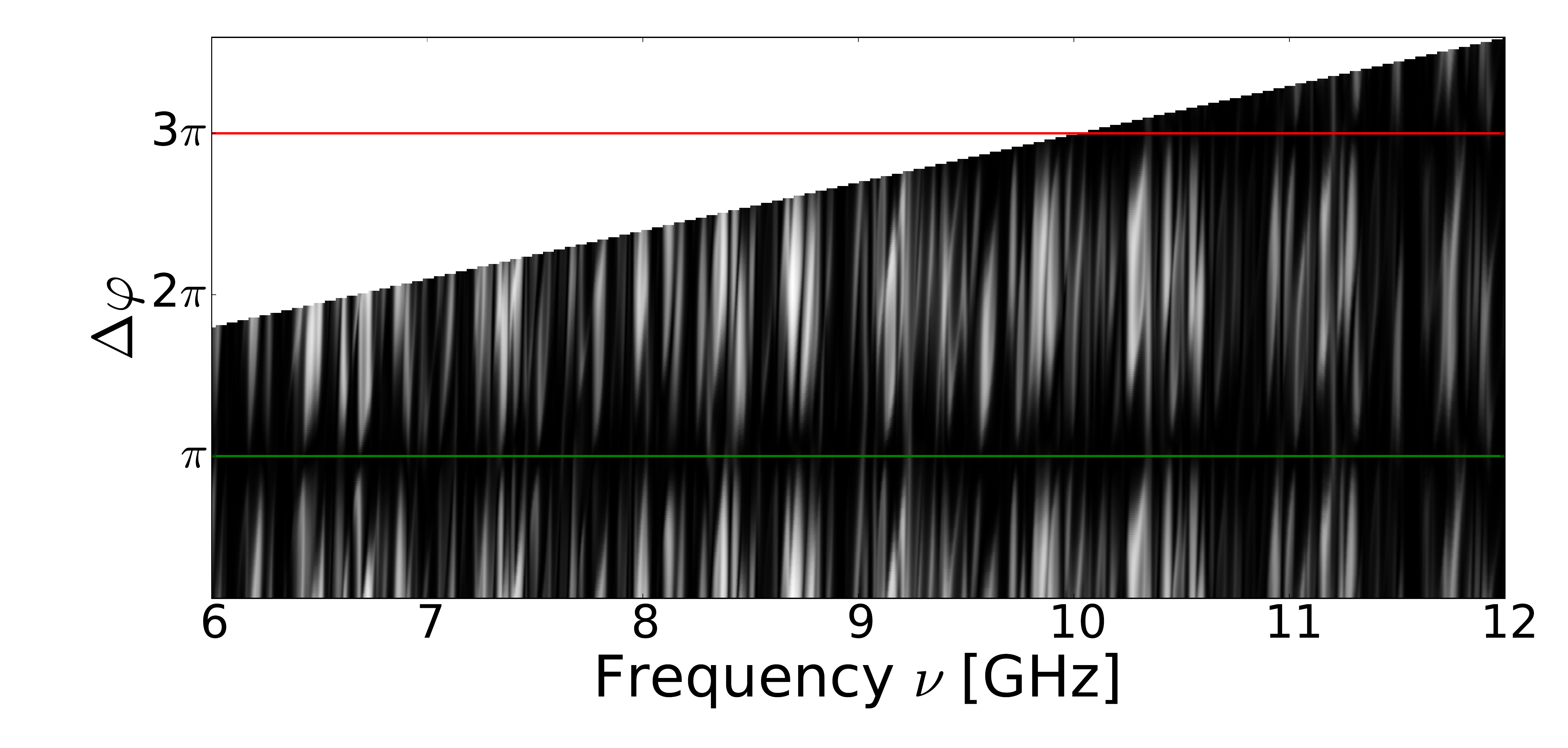}
  \caption{\label{fig:S12_spectra}
  (Top) Transmission $\left|S_{12}\right|^2$ in dependence of frequency for constant $\Delta l$ in a gray scale, where black corresponds to zero and white to maximal transmission. The measurements for different $\Delta l$ are stacked onto each other. (Bottom) The same data but rearranged to constant $\Delta\varphi$.}
\end{figure}

Figure~\ref{fig:graph}(b) shows the schematic drawing of one realized graph and Fig.~\ref{fig:graph}(c) the photograph of the corresponding experimental realization. The bonds of the graphs were formed by Huber~\&~Suhner EZ-141 coaxial semirigid cables with SMA connectors coupled by $T$ junctions at the nodes. The phase shifters (ATM, P1507) were equipped with motors to allow for an automatic stepping. Reflection and transmission measurements were performed with an Agilent 8720ES vector network analyzer (VNA) with the two ports $P_1$ and $P_2$ at equivalent positions of the two subgraphs. The corresponding reflection and transmission amplitudes will be denoted in the following by $S_{ij}$, where $i,j=1,2$ is defined by the port. The operating range of the circulators (Aerotek I70-1FFF) positioned at nodes $7$ and $\bar{7}$ extended from 6 to 12\,GHz. Therefore, the {analysis of the data was restricted to this window.

We started by taking a series of measurements for constant $\Delta l$. Figure~\ref{fig:S12_spectra}(top) shows the transmission for altogether 396 $\Delta l$ values stacked onto each other between $\Delta l_{\mathrm{min}}\approx 0$ and $\Delta l_{\mathrm{max}}= 4.4$\,cm in a gray scale. The lines for $\Delta\varphi=\pi$ and $\Delta\varphi=3\pi$ are marked in red and green, respectively. Next, a variable transformation from $\Delta l$ to $\Delta\varphi$ was performed using Eq.~(\ref{eq:phase}) to obtain the transmission $S_{12}$ for constant $\Delta\varphi$. The result is shown in Fig.~\ref{fig:S12_spectra}(bottom).

For a given frequency $\nu$, the maximum $\Delta\varphi$ accessible is, according to Eq.~(\ref{eq:phase}), given by $\Delta\varphi_{\mathbf{max}}=(2\pi\Delta l_{\mathbf{max}}/c)\nu$. The inaccessible regime above this limit is left white in Fig.~\ref{fig:S12_spectra}(bottom). As expected, the pattern is periodic in $\Delta\varphi$ with period $2\pi$. For $\Delta\varphi=\pi$ and $\Delta\varphi=3\pi$, the transmission is strongly suppressed. This is an interference effect: All transmission paths from $P_1$ to $P_2$ come in pairs, e.g., the paths $0732\bar{1}\bar{6}\bar{5}\bar{0}$ and $0561\bar{2}\bar{3}\bar{7}\bar{0}$; see Fig.~\ref{fig:graph}(b). One of these passes through one phase shifter, whereas its partner passes through the other, and as a result, their lengths differ by $\Delta l$. Depending on the resulting $\Delta\varphi$, this gives rise to constructive or destructive interference.

\begin{figure}
  \includegraphics[width=\columnwidth]{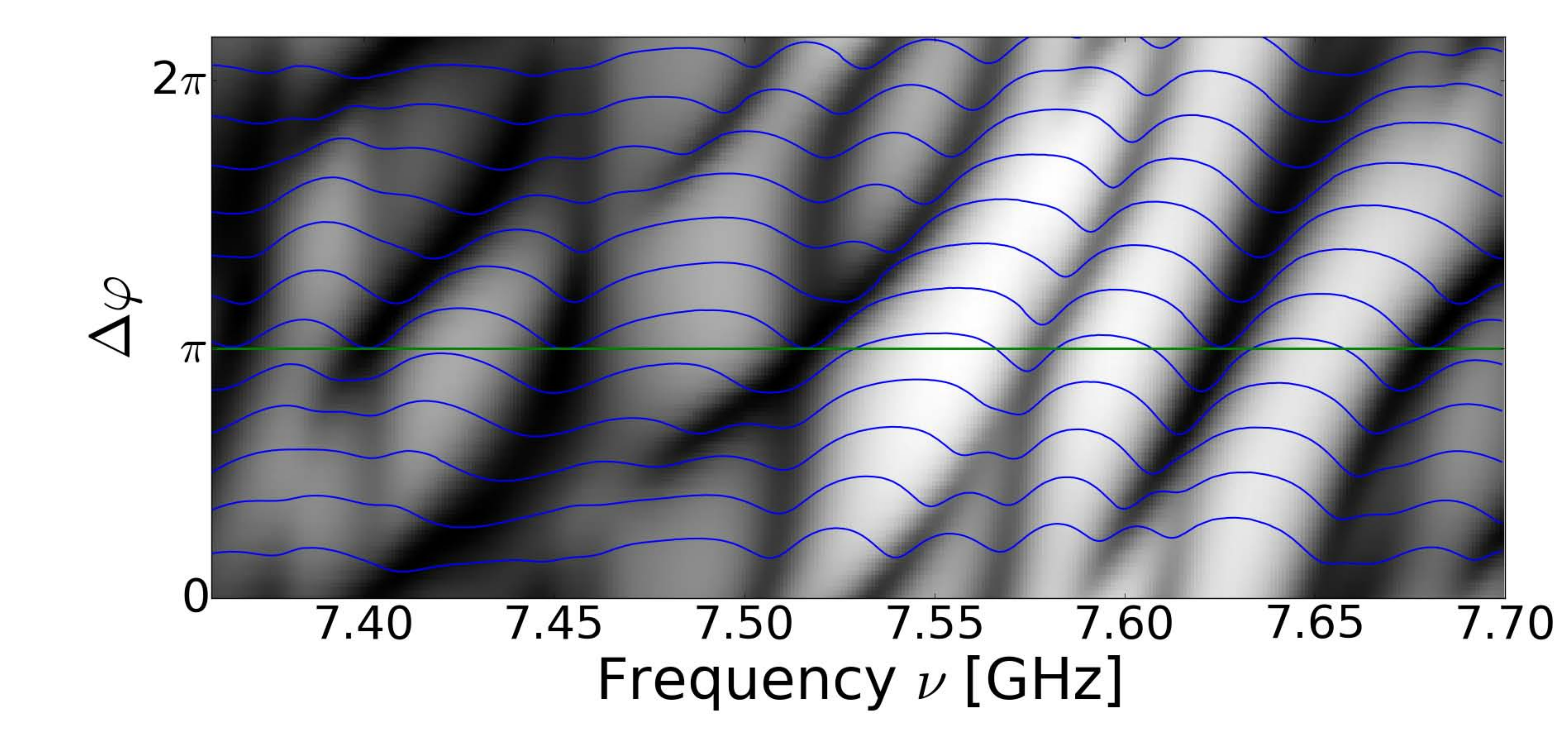}
  \caption{\label{fig:S11_spectra}
  Reflection $\left|S_{11}\right|^2$ in dependence of frequency in a shade plot. The results for different $\Delta\varphi$ are stacked onto each other.}
\end{figure}

Because of the lack of transmission at $\Delta\varphi=\pi$, we proceeded to analyze the reflection $\left|S_{11}\right|^2$. This is shown in Fig.~\ref{fig:S11_spectra} for a small frequency window and for different $\Delta\varphi$, again stacked on top of each other in a shaded plot. Each eigenfrequency shows up as a dip. One clearly observes the formation of Kramers doublets at the $\pi$ line, and their splitting into singlets when departing from this line. There is a complete equivalence to the Zeeman splitting of spin doublets: In the present experiment, the antiunitary symmetry is destroyed when departing from the $\pi$ line, whereas for conventional spin systems, this effect occurs when applying a magnetic field. This is a clear confirmation that we were successful in constructing a graph with antiunitary symmetry $T$ with $T^2=-1$. The distances of the six Kramers doublets seen in Fig.~\ref{fig:S11_spectra} at the $\pi$ line are equal within 20\,\%. This shows a clear tendency of the levels towards an equal level spacing at the $\pi$ line, one of the fingerprints for a GSE spectrum.

\begin{figure}[b]
  \includegraphics[width=.95\columnwidth]{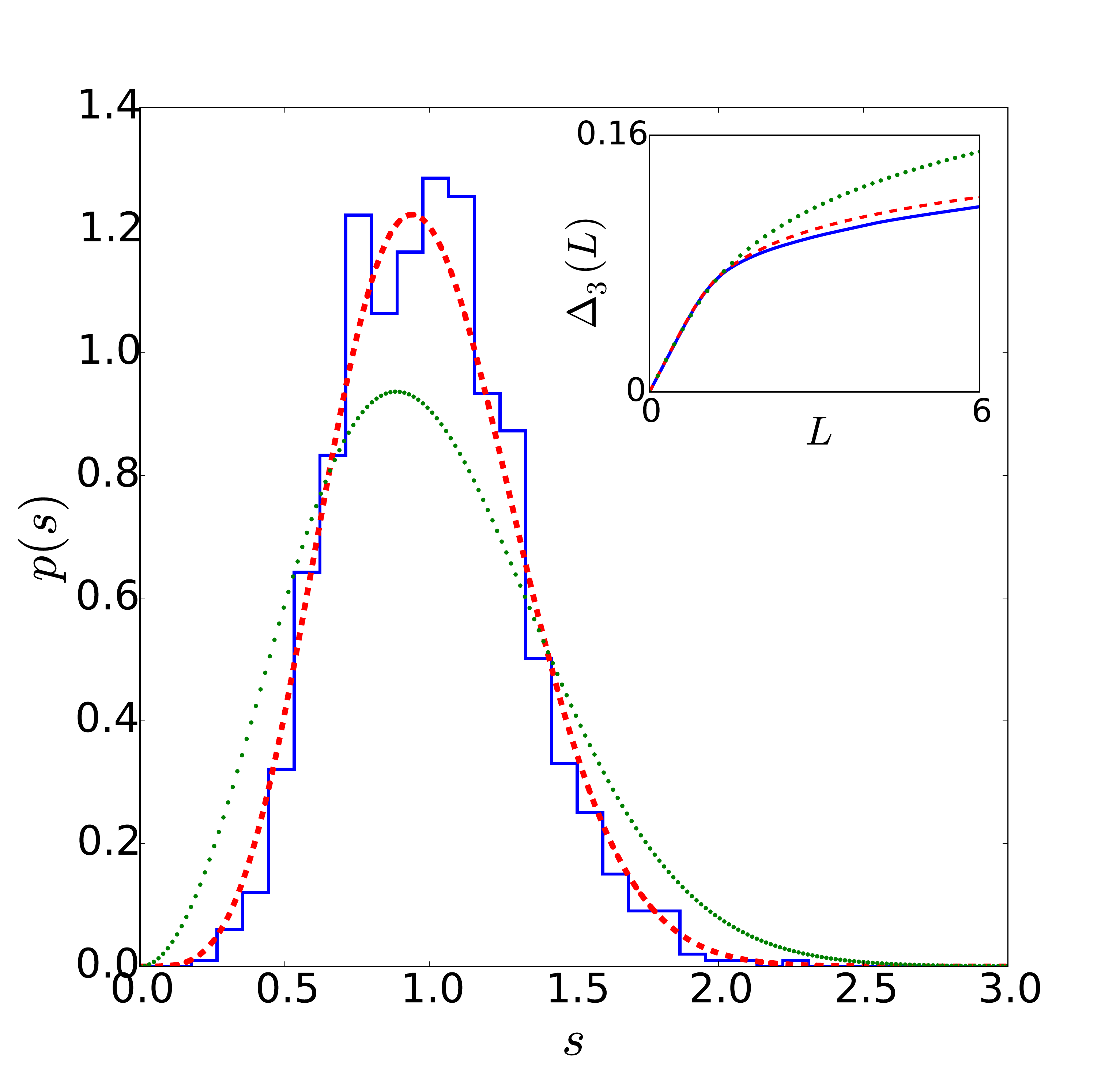}
  \caption{\label{fig:levspac} Spectral nearest neighbor distance distribution obtained by superimposing the results from eight different spectra (blue). The dashed red and dotted green lines correspond to GSE and GUE Wigner distributions, respectively; see Eqs.~(\ref{eq:GSE}) and (\ref{eq:GUE}). The inset shows the spectral rigidity for the same data set (blue), again with random matrix predictions for the GSE and the GUE in red and green, respectively.}
\end{figure}

To obtain the complete eigenfrequency spectrum, we proceeded as follows: Though there are two coupled channels, they are equivalent to one effective single channel for $\Delta\varphi=\pi$ due to symmetry. In this case, the scattering matrix reduces to a phase factor $S=e^{i\varphi}=(1-iK)/(1+iK)$, i.e., $\varphi=-2\arctan(K)$. $K$ may be written as a sum over resonance poles $a_n/(x-x_n)$, meaning stepwise phase changes at $x=x_n$. By taking the phase derivative, these steps turn into sharp peaks with widths limited by absorption (which were discarded in the argumentation). This allowed for an automatic determination of about 90\,\% of the eigenvalues. With the additional information from the spectral level dynamics (see Fig.~\ref{fig:S11_spectra}), the missing ones could be easily identified. About 10\,\% of the Kramers doublets split due to experimental imperfections. Whenever this was evident from the level dynamics, the resulting two resonances were replaced by a single one in the middle.

The integrated density of eigenfrequencies may be written as $n(\nu)=n_\mathrm{Weyl}(\nu)+ n_\mathrm{fluc}(\nu)$, where the average part is given by Weyl's law $n_\mathrm{Weyl}(\nu)=(\pi/L)(2\pi\nu/c)$, with $L$ denoting the sum of all bond lengths \cite{kot99a}. The fluctuating part $n_\mathrm{fluc}(\nu)$ reflects the influence of the periodic orbits \cite{gut90}. We determined $n_\mathrm{fluc}(\nu)$ by fitting a straight line to the experimental integrated density of eigenfrequencies and subtracting the linear part. A small number of missing or misidentified resonances showing up in stepwise changes of $n_\mathrm{fluc}(\nu)$ enabled the final correction of the spectrum. From the fit, the length was obtained, e.\,g., $L= 2.93$\,m for the graph shown in Fig.~\ref{fig:graph}(c). The nearest neighbor spacings $s$ are calculated as the difference $s_n=n_\mathrm{Weyl}(\nu_{n+1})-n_\mathrm{Weyl}(\nu_{n})$ for the individual graph guaranteeing a mean level spacing $\langle s \rangle=1$.

Figure~\ref{fig:levspac} shows the resulting distribution of spacings between neighboring levels in units of the mean level spacing. To improve the statistics, the results from eight different graphs were superimposed, leading to 1006 Kramers doublets. The red solid and the green dotted line correspond to the Wigner prediction for the GSE
\begin{equation}\label{eq:GSE}
  p_\mathrm{GSE}(s)=\frac{2^{18}}{3^6\pi^3}s^4\exp\left(-\frac{64}{9\pi}s^2\right)
\end{equation}
and the GUE
\begin{equation}\label{eq:GUE}
  p_{\,\mathrm{GUE}}(s)=\frac{32}{\pi^2}s^2\exp\left(-\frac{4}{\pi}s^2\right)\,,
\end{equation}
respectively. The experimental result fits well to the GSE distribution, and though the statistical evidence as yet is only moderate, it is clearly at odds with a GUE distribution. The inset of Fig.~\ref{fig:levspac} shows the associated spectral rigidity $\Delta_3(L)$ \cite{meh91}. Again, a good agreement with the GSE random matrix prediction is found. The small deviations suggest some percents of misidentified levels, which would have only a minor influence on the nearest neighbor spacings distribution but would distort long-range correlations.

It needed half a century after the establishment of random matrix theory by Wigner, Dyson, Mehta, and others to arrive at the present experimental realization of the third of the three classical random matrix ensembles. It might be considered surprising that two bonds between the two subgraphs are already sufficient to turn the two GUE spectra of the disconnected
subgraphs into a GSE spectrum for the total graph. On the other hand, the present statistical evidence is not yet sufficient to determine whether more connecting bonds are needed in order to reduce the minor differences in the level spacing statistics. Further studies are, thus, required. The dependence of the level dynamics on $\Delta\varphi$ offers a promising research direction due to the interesting feature that all three classical ensembles are present, namely, the GSE for $\Delta\varphi=\pi$, the GOE for $\Delta\varphi=0$, and the GUE in between. However, the most promising future aspect is undoubtedly that the whole spin 1/2 physics \cite{dub13} is now accessible to microwave analogue studies.

\begin{acknowledgments}
This work was funded by the Deutsche Forschungsgemeinschaft via the individual Grant No. STO 157/16-1 and No. KU 1525/3-1. C.\,H.\,J.~acknowledges the
Leverhulme Trust (Grant No. ECF-2014-448) for financial support.
\end{acknowledgments}

\end{document}